\documentstyle[psfig,twoside,fleqn,espcrc2]{article}

\newcommand{\AmS}{{\protect\the\textfont2
  A\kern-.1667em\lower.5ex\hbox{M}\kern-.125emS}}

\title{Absorption dips in GRO\,J1655$-$40: mapping the inner accretion disk}

\author{E.~Kuulkers\address{Astrophysics, University of Oxford, Nuclear and
       Astrophysics Laboratory, Keble Road, Oxford OX1 3RH, United
       Kingdom},
       R.~Wijnands\address{Astronomical Institute ``Anton Pannekoek'',
       University of Amsterdam and Center for High-Energy Astrophysics,
       Kruislaan 403, NL-1098 SJ Amsterdam, the Netherlands}, 
       T.~Belloni$^{\rm b}$, 
       M.~M\'endez$^{\rm b,}$\address{Facultad de Ciencias
       Astron\'omicas y Geof\'{\i}sicas, Universidad Nacional de La Plata, 
       Paseo del Bosque S/N, 1900 La Plata, Argentina}, 
       M.~van der Klis$^{\rm b}$ \&\ 
       J.~van Paradijs$^{\rm b,}$\address{Physics 
       Department, University of Alabama in Huntsville, Huntsville, AL 35899, 
       USA}}
       
\begin{document}

\begin{abstract}
Using the RXTE PCA we discovered dips in the X-ray light curves 
of the black-hole candidate GRO\,J1655$-$40 during outburst. They are
short ($\sim$minute) and deep (down to $\sim$8\%\ of the out-of-dip intensity).
Similar kind of dips were found in 90\,s measurements of the RXTE ASM 
during the same outburst. The occurrences of the dips 
are consistent with the optically determined orbital period, and were found 
between photometric orbital phases 0.72 and 0.86. This constitutes the first 
evidence for orbital variations in X-rays for GRO\,J1655$-$40. 
The PCA data indicate that an absorbing medium is 
responsible for these dips. Using these results we are able 
constrain the extent of the absorbing medium and the central X-ray source.
GRO\,J1655$-$40 was in the canonical high state during our PCA observations.
\end{abstract}

\maketitle

\section{Introduction}

The low-mass X-ray binary and soft X-ray transient GRO\,J1655$-$40 was 
discovered during an outburst in 1994 and since then has shown irregular outburst 
activity (e.g., Zhang et al.\ \cite{zes97}). Radio observations during its
1994 outburst revealed jets moving at relativistic (apparently superluminal) 
speeds (Tingay et al.\ \cite{txx95}, Hjellming \&\ Rupen \cite{hr95}).
Dynamical measurements suggest that the compact star in GRO\,J1655$-$40 
is a black hole, with a mass of $\sim$7\,M$_{\odot}$, while the companion star
has a mass of $\sim$2\,M$_{\odot}$;
its orbital period is 2.62~days and its inclination is $\sim$70$^{\circ}$
(Orosz \&\ Bailyn \cite{ob97}, van der Hooft et al.\ \cite{hha97}). 

X-ray intensity dips caused by an intervening medium have now been 
found in the light curves of various low-mass and high-mass X-ray binaries
(e.g., Parmar \&\ White \cite{pw88}, Marshall et al.\ \cite{mmp93}, 
Saraswat et al.\ \cite{sym96}, and references therein).
During the majority of these dips the X-ray spectra harden, which is 
indicative of photo-electric absorption of radiation from the central
X-ray source. However, a simple neutral and uniform medium which absorbs the 
emission does not fit the X-ray spectra. Instead, the spectra reveal
an excess flux at low energies (typically $<$4\,keV) compared to that 
expected from the amount of absorption estimated from data above $\sim$4\,keV.

\begin{figure*}[htb]
\centerline{
\psfig{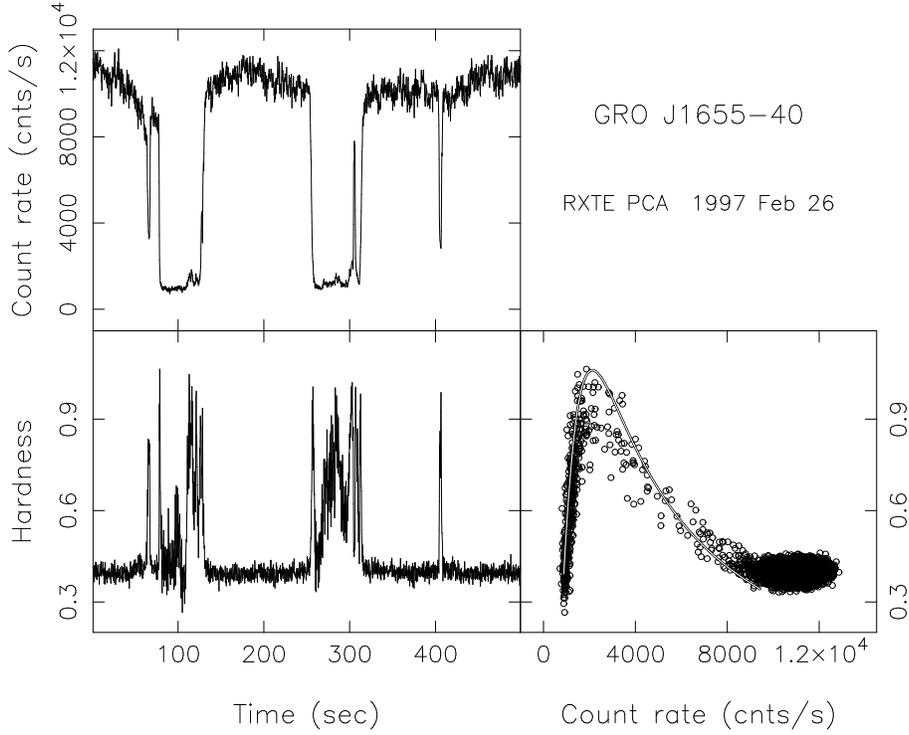}
}
\caption{The light curve (upper left panel), hardness curve (lower left panel) 
and hardness-intensity diagram (HID, lower right panel) of GRO\,J1655$-$40. 
The hardness is defined as the ratio of 
the count rates in the 5.0--13.0\,keV and 2.0--5.0\,keV bands. The time 
resolution in the left panel is 0.25\,s. 
T=0\,s corresponds to 1997 Feb 26, 21:22:21~UTC.
The data points in the HID (0.25\,s averages) are given by open circles.
The line through these data points corresponds to a simple model of the 
X-ray spectral behaviour as described in the text.}
\end{figure*}

In this contribution we report on such dips in the 
X-ray light curves of GRO\,J1655$-$40 seen with the 
{\it Rossi X-ray Timing Explorer} (RXTE) during its 1996/1997 outburst.
A full report can be found in Kuulkers et al.\ (\cite{kwb98}).

\section{Observations}

\subsection{Periodic drops in the intensity}

The first public RXTE Target of Opportunity Obser\-vations of 
GRO\,J1655$-$40 were 
performed on 1997 February 26 19:34--23:30~UTC ($\sim$JD\,2450506). 
The {\it Proportional Counter Array} (PCA) data were collected with a time 
resolution of 16\,s 
(129 photon energy channels, covering 2.0--60\,keV) and 125\,$\mu$s 
(3 energy channels, covering 2.0--5.0--8.7--13.0\,keV). The 2.0--13.0\,keV 
count rate of GRO\,J1655$-$40 was generally $\sim$11\,500\,cts\,s$^{-1}$. 
However, 2 sharp deep drops down to $\sim$1000\,cts\,s$^{-1}$ lasting 
$\sim$50--60\,s, preceded and followed by short dips, occurred
(Fig.~1, upper panel).
The fall time scales of the main dips were 2--4\,s, while the rise time scales 
were 3--5\,s. 

The {\it All Sky Monitor} (ASM) onboard RXTE scans the sky in series of 90\,s 
dwells in three energy bands, 
1.5--3, 3--5, and 5--12\,keV. Due to satellite motion and a $\sim$40\%\ duty 
cycle, any given source is scanned 5--10 times per day. In Fig.~2 we show 
the 1996/1997 outburst light curve (2-12\,keV) provided by the RXTE ASM team, 
covering the period from 1996 February 21 to 1997 October 3.
Twelve clear drops in the intensity can be seen down by $\sim$25--95\%, which are
indicated by arrows. During these dips the spectral hardness increases
(Fig.~1, lower left panel). 

\begin{figure*}[htb]
\centerline{
\psfig{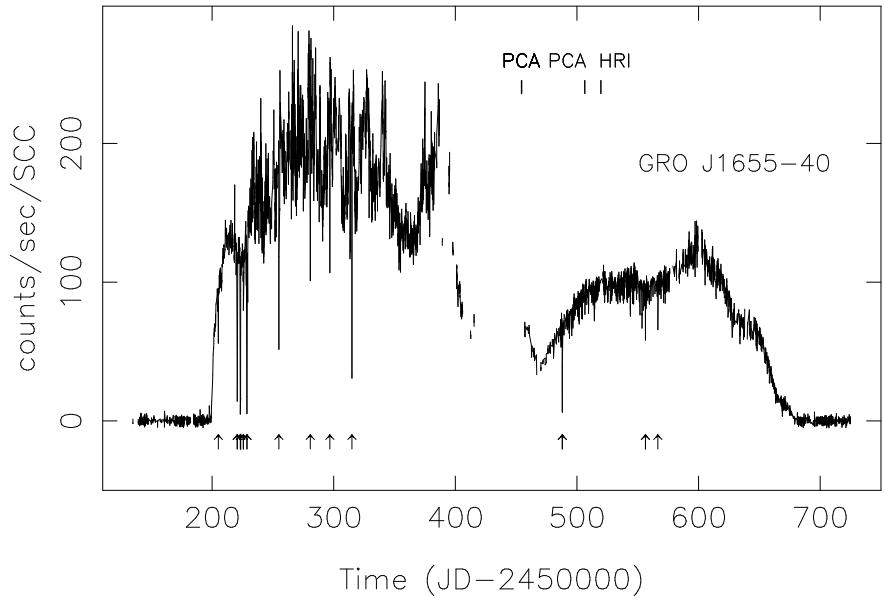}
}
\caption{RXTE ASM lightcurve of GRO\,J1655$-$40 of data 
from individual dwells of $\sim$90\,s from 1996 Feb 21 (JD\,2\,450\,135) to 
1997 Oct 3 (JD\,2\,450\,724). Datapoints separated by $<$2\,d have been 
connected to guide the eye. Clearly, deep sharp dips can be seen, which are
indicated by arrows. Indicated are also the times of the dips observed with 
the RXTE PCA and the ROSAT HRI.}
\end{figure*}

The occurrences of the ASM and PCA dips are best fit with a period of 
2.6213$\pm$0.0005 days (1$\sigma$).
Recently, we became aware of more dips observed by the PCA 
($\sim$JD\,2450488; R.~Remillard, 1997, private communication) and the 
ROSAT HRI ($\sim$JD\,2450520; Kuulkers et al.\ \cite{kxx97}); the dips
observed by the PCA and the ROSAT HRI 
are also indicated in Fig.~2. All the observed dips occurred between 
photometric orbital phases (Orosz \&\ Bailyn \cite{ob97}, van der Hooft et al.\
\cite{hha97}) 0.72 and 0.86. As an example, in Fig.~3 we show a small part of 
the ASM data during which dips could be seen in every orbital cycle, folded on 
the orbital ephemeris (Orosz \&\ Bailyn \cite{ob97}, van der Hooft et al.\
\cite{hha97}).

\begin{figure*}[htb]
\centerline{
\psfig{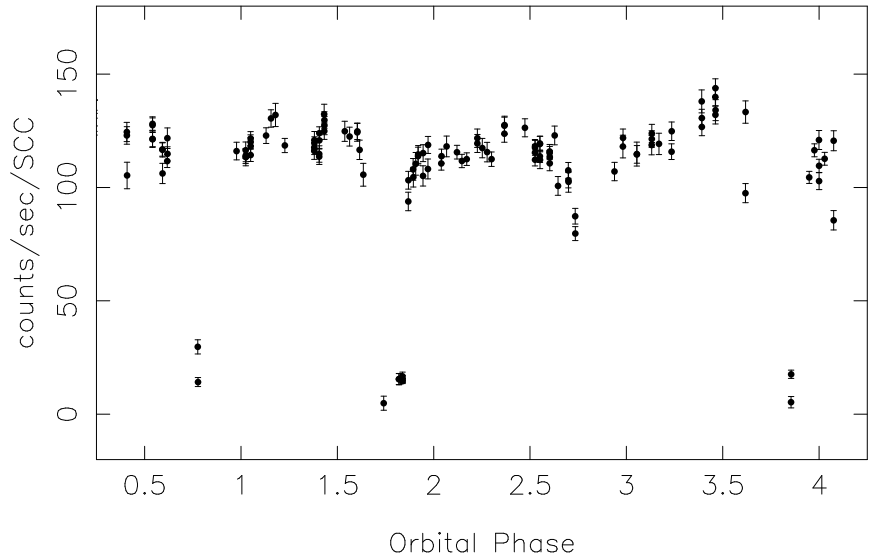}
}
\caption{RXTE ASM lightcurve of GRO\,J1655$-$40 using data of individual 
90\,s dwells as a function of orbital phase. The light curve spans from 1996 
May 15--25 (JD\,2450218--JD\,2450228).}
\end{figure*}

\subsection{Absorption}

The PCA hardness curve (Fig.~1, lower left panel) and 
corresponding hardness vs.\ intensity diagram 
(HID, Fig.~1, lower right panel) show that as the 
intensity drops the X-ray spectrum becomes much harder, until a certain 
threshold count rate of $\sim$2000\,cts\,s$^{-1}$ is reached.
During the last part of the fall the spectrum softens again to almost the 
same value as the out-of-dip level. The reverse behaviour is seen during the 
rise of the dips. In the HID the source always follows the same loop, also 
during the small pre- and after-dips and during the short spike 
in the second dip.
The spectral hardening between the out-of-dip count rate and the 
threshold value suggests that absorption is involved in the process giving 
rise to the dips. 

We fitted both the out-of-dip (persistent) and the dips energy spectra.
For the fits to the out-of-dip spectrum we used the model usually employed for 
black-hole candidates, i.e., a disk-black body (DBB) plus a power law.
This resulted in: 
N$_{\rm H}=(2.17 \pm 0.11) 10^{22}$\,cm$^{-2}$, for DBB parameters 
T$_{\rm in}$ and R$_{\rm in}$ $1.110 \pm 0.002$\,keV and $24.7 \pm 1.0$\,km,
respectively, and the power-law parameters $\Gamma$ and normalization 
A$_{\rm pl}$ $2.4 \pm 0.02$ and 
$0.29 \pm 0.01$\,cm$^{-2}$\,s$^{-1}$\,keV$^{-1}$, respectively
(reduced $\chi^2$ of 1.6 for 54 degrees of freedom [dof]). The out-of-dip
X-ray flux (2--30\,keV) was
$\sim$2.5$\times$10$^{-8}$\,erg\,cm$^{-2}$\,s$^{-1}$.
This shows that the spectrum was soft.

The dip spectra were not well fitted by homogeneous absorption of the 
out-of-dip DBB and power-law components by cold material, especially at low 
energies. The observed flux below 
$\sim$6\,keV is much in excess from that expected in this model.
We performed simple fits to the spectra by modelling 
this `low energy excess' either as a separate component 
(power law or black body), or by partial covering absorption of the persistent 
components (e.g.\ Marshall et al.\ \cite{mmp93}).
In all the dip spectral fits the persistent power-law component was found to 
be absent. The low-energy excess 
contributes only 6--7\%\ of the out-of-dip flux. Depending on the model we 
found that the absorption of the DBB component increased up to 
$\sim$25--200$\times$10$^{22}$ at the lowest mean dip intensities,
see Kuulkers et al.\ (\cite{kwb98}).

To see if we can qualitatively reproduce the observed shapes of the 
HID, we calculated several sequences of X-ray spectra and determined
intensity and hardness values. 
The out-of-dip spectrum was modeled by the persistent DBB component 
(only subject to interstellar absorption) as given above, plus the 
low-energy excess contribution modeled by a black body (see Kuulkers et al.\
(\cite{kwb98}). In the dip we linearly increased the absorption of the 
persistent DBB component from zero up to 150$\times$10$^{22}$\,cm$^{-2}$, 
fixing the rest of the 
parameters to those given by the out-of-dip spectrum.
The results are plotted as a line in the lower right panel of Fig.~1, 
and show that a gradual increase in 
absorption of the DBB component can reproduce the observed dip behaviour.

\subsection{Fast timing behaviour}

In order to investigate the state of the source (see e.g.\ van der Klis 
\cite{k95}) we computed power density spectra (PDS) for 256\,s data 
stretches for the out-of-dip light curves. The mean PDS is shown in Fig.~4,
and is well fitted with a single power law (reduced $\chi^2$ of 202 for 186
dof). The mean fractional rms variations
(0.01--1\,Hz, 2--13\,keV) and power-law index are 2.80$\pm$0.06\%\ and 
0.98$\pm$0.01, respectively. We found no energy dependence of the PDS shapes.
We also computed PDS during the dips; then we used data stretches of 16\,s. 
No clear noise component could be seen in the power spectra during the dip, with
a 2$\sigma$ upper limit of 4.6\%\ and 3.6\%\ (0.01--100\,Hz, 2--13\,keV),
assuming a power-law shaped noise component with index 1.
The properties of the out-of-dip PDS and the strong soft DBB component 
together with a steep power law in the X-ray energy spectra indicate that 
when the dips occurred GRO\,J1655$-$44 was in the so-called high state.

\section{Discussion}

\subsection{Absorption dips}

We have discovered short-term ($\sim$minutes) 
X-ray deep dipping behaviour of GRO\,J1655$-$40 in RXTE PCA and ASM data.
The best fit period of the occurrence of these dips is consistent with the 
optical period of the system (2.62168$\pm$0.00014~days; van der Hooft et al.\
\cite{hha97}; see also Orosz \&\ Bailyn \cite{ob97}). This therefore 
constitutes the first evidence of the orbital period in GRO\,J1655$-$40 in 
X-rays. All these dips occurred between photometric orbital phases 0.72 and 
0.86. 

The phasing of the occurrence of the X-ray dips is very similar to that 
observed in the low-mass X-ray binary dip sources (e.g., Parmar \&\ White 
\cite{pw88}). The inclination of such sources which show only dips and no 
eclipses are in the range 60--75$^{\circ}$ (e.g., Frank et al., \cite{fkl87}). 
The inclination inferred for GRO\,J1655$-$40 ($\sim$70$^{\circ}$, Orosz \&\ 
Bailyn \cite{ob97}, van der Hooft et al.\ \cite{hha97}) is in agreement with 
this. This suggests a similar origin for the cause of the dips in 
GRO\,J1655$-$40 and the low-mass X-ray binary dip sources.
Interaction of the inflowing gas stream from the secondary with the 
outer edge of the disk may cause a thickening of the outer edge and/or
material above or below the disk in the expected 0.6--0.0 phase range
(e.g., Parmar \&\ White \cite{pw88}) and may well cause the observed dipping
behaviour. Since the dip durations are rather short compared to the 
orbital period, we expect that the dips in GRO\,J1655$-$40 are caused by 
filaments of matter 
above or below the disk originating from the stream-disk impact region.

\begin{figure}[htb]
\psfig{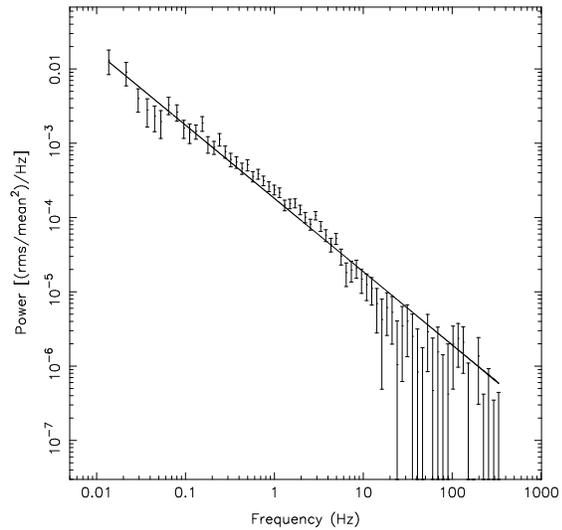}
\caption{Mean power density spectrum of the out-of-dip light curve
(2--13\,keV). The corresponding power-law fit is shown, with mean fractional 
rms variation (0.01--1\,Hz) of $\sim$3\%\ and power-law index of $\sim$1.}
\end{figure}

\subsection{An atlas of GRO\,J1655$-$40}

For GRO\,J1655$-$40 the system parameters have been well determined
(Orosz \&\ Bailyn \cite{ob97}, Van der Hooft et al.\ \cite{hha97}). In this 
system the time scales of the dips imposes constraints on the sizes
of the different emitting and absorbing media.
The fall and rise time ($t_{\rm r,f}$$\sim$3.5\,s) 
constrain the size of the region which is "obscured";
in fact, this gives an upper limit on the size, because the region over 
which the column density increases significantly also has a finite width
(Leahy, Yoshida, \&\ Matsuoka \cite{lym94}). 
The duration of the dips ($t_{\rm dip}$$\sim$55\,sec) 
constrains the size of the absorbing medium itself.

Since $t_{\rm r,f}$$\ll$$t_{\rm dip}$, we may assume that 
the absorbed X-ray source is much smaller than the absorbing medium. 
A medium which crosses a 
point-like central X-ray source may produce irregular X-ray dips, whereas 
crossing an extended region such as an accretion disk corona may produce 
smooth and
longer energy independent modulations (e.g., Parmar \&\ White \cite{pw88}). 
If the medium corotates in the binary 
frame and is located at a radius which is smaller than the outer disk radius
($r_{\rm d}$$\sim$0.85\,R$_{\rm L}$ [Orosz \&\ Bailyn \cite{ob97}], where 
R$_{\rm L}$ is the effective Roche lobe radius of the black hole) the upper
limit on the size of the X-ray emitting region is $\sim$460\,km. If the 
medium corotates with matter in the accretion disk (i.e., with a Kepler 
velocity), the upper limit becomes $\sim$2000\,km. Similar reasoning 
gives an approximate upper limit on the size of the absorbing medium of 
$\sim$7200\,km or $\sim$32\,000\,km, in the case of rotation within the binary 
frame or corotation in the accretion disk.

We note that the partial covering absorption model to fit the spectra during the
dips may impose physical problems. In such a model the absorbing medium consists
of blobs of material and an individual blob only partially covers the central 
X-ray source. This means that the individual blobs must be substantially smaller
than the inferred size of the X-ray emitting region. Since the size of the 
absorbing medium is considerably (15--70 times) larger than the X-ray 
emitting region there must be many of such blobs in the absorbing medium.
Then one would expect many of them along the line of sight, and that they
wouldd average out to a more or less uniform density medium, unless the 
radial extent of the absorbing medium is much smaller than the azimuthal extent.

\subsection{4U\,1630$-$47 and GRS\,1915+105}

Recently, a dip in the light curve of 4U\,1630$-$47 during its 1996 outburst
has been found in RXTE PCA data (Kuulkers et al.\ \cite{kwb98}; 
Tomsick et al.\ \cite{tlk98}). Kuulkers et al.\ (\cite{kkp97}; \cite{kpk97}) 
pointed out similarities in the X-ray behaviour between 4U\,1630$-$47 and 
GRO\,J1655$-$40, and postulated that they are similar systems.
The nature of the compact star in 4U\,1630$-$47, however, is unknown.
Its X-ray spectral (e.g.\ Barret et al.\ \cite{bmg96})
and X-ray timing (Kuulkers et al.\ \cite{kkp97}) properties 
during outburst suggest it is a black-hole. The dip spectral behaviour of
4U\,1630$-$47 is similar to that seen in GRO\,J1655$-$40; it 
was therefore proposed that 4U\,1630$-$47 is also seen at a relatively high 
inclination, i.e.\ 60--75$^{\circ}$ (Kuulkers et al.\ \cite{kwb98}). 
We note that the `dipping' behaviour seen in the other Galactic superluminal 
jet source, GRS\,1915+105, is much more complex than in 
GRO\,J1655$-$40 and 4U\,1630$-$47. This dipping behaviour has been proposed to 
be due to thermal-viscous 
instabilities in the inner disk (Belloni et al.\ \cite{bmk97}) and is therefore
not due to absorption events.

\section*{Acknowledgements}

This work was supported in part by the Netherlands Organisation for
Scientific Research (NWO) under grant PGS 78-277 and by the
Netherlands Foundation for research in astronomy (ASTRON).  MM is a
fellow of the Consejo Nacional de Investigaciones Cient\'{\i}ficas y
T\'ecnicas de la Rep\'ublica Argentina. JvP acknowledges support from the 
National Aeronautics and Space Administration through contract NAG5-3269.
The RXTE ASM data used are quick-look results provided by the ASM/RXTE team.

\end{document}